# 4$f$-hybridization strength in Ce$_m$M$_n$In$_{3m+2n}$ heavy-fermion compounds studied by Angle-Resolved Photoemission Spectroscopy

Jiao-Jiao Song,[1] Yang Luo,[1] Chen Zhang,[1] Qi-Yi Wu,[1] Tomasz Durakiewicz,[2] Yasmine Sassa,[3,4] Oscar Tjernberg,[5] Martin Månsson,[5] Magnus H. Berntsen,[5] Yin-Zou Zhao,[1] Hao Liu,[1] Shuang-Xing Zhu,[1] Zi-Teng Liu,[1] Fan-Ying Wu,[1] Shu-Yu Liu,[1] Eric D. Bauer,[6] Ján Rusz,[3] Peter M. Oppeneer,[3] Ya-Hua Yuan,[1] Yu-Xia Duan,[1] and Jian-Qiao Meng[1, *]

[1]*School of Physics and Electronics, Central South University, Changsha 410083, Hunan, China*
[2]*Institute of Physics, Maria Curie Sklodowska University, 20-031 Lublin, Poland*
[3]*Department of Physics and Astronomy, Uppsala University, Box 516, S-75120 Uppsala, Sweden*
[4]*Department of Physics, Chalmers University of Technology, 41296 Göteborg, Sweden*
[5]*Department of Applied Physics, KTH Royal Institute of Technology, AlbaNova Universitetscentrum, 106 91 Stockholm, Sweden*
[6]*Condensed Matter and Magnet Science Group, Los Alamos National Laboratory, Los Alamos, New Mexico 87545, USA*
(Dated: Monday 16$^{\text{th}}$ August, 2021)

We systemically investigate the nature of Ce 4$f$ electrons in structurally layered heavy-fermion compounds Ce$_m$M$_n$In$_{3m+2n}$ (with $M$=Co, Rh, Ir, and Pt, $m$=1, 2, $n$=0 - 2), at low temperature using on-resonance angle-resolved photoemission spectroscopy. Three heavy quasiparticle bands $f^0$, $f^1_{7/2}$ and $f^1_{5/2}$, are observed in all compounds, but their intensities and energy locations vary greatly with materials. The strong $f^0$ states imply that the localized electron behavior dominates the Ce 4$f$ states. The Ce 4$f$ electrons are partially hybridized with the conduction electrons, making them have the dual nature of localization and itinerant. Our quantitative comparison reveals that the $f^1_{5/2}$-$f^0$ intensity ratio is more suitable to reflect the 4$f$-state hybridization strength.

## INTRODUCTION

The heavy-fermion (HF) system is an excellent platform for studying the behavior of correlated electrons in unconventional superconductors, quantum criticality and non-Fermi Liquid, antiferromagnetism [1–3]. Previous studies on HF superconductors reveal that both dimensionality [4–6] and hybridization [7] are closely related to superconductivity. Tuning dimensionality and hybridization is the key to revealing the microscopic mechanism of unconventional superconductivity in heavy fermions.

It is found that the lattice parameters ratio $c/a$ of CeMIn$_5$ ($M$ = Co, Rh, Ir) has a roughly linear relationship with its superconducting transition temperature $T_c$ [8]. Pu-based 115 materials $T_c$ is also linear in $c/a$ [9]. Uniaxial pressure measurement on CeIrIn$_5$ revealed that $T_c$ changed linearly with both $a$-axis and $c$-axis pressure [10]. The $c/a$ ratio may be the control parameter for the scaling of the superconducting dome width and $T_c^{max}$. The value of $T_c$ in magnetically mediated superconductors is believed to be dependent on dimensionality [4–6]. Superconducting properties can be enhanced by lowering dimensionality. It is suggested that the difference in hybridization sets the overall temperature scale for each family, but dimensionality governs the behavior within each family.

To explore the character of $f$ electrons is essential to understanding the complex physics in HF compounds. $f$-electrons participate in the formation of Fermi surface through hybridization with conduction electrons. The localized or itinerant properties of $f$-electrons, i.e., hybridization between conduction and $f$ electrons ($c$-$f$), will significantly affect the band structure near the Fermi energy ($E_F$). After $c$-$f$ electron hybridization, the complex itinerant heavy electron state is formed, which leads to the delocalization of $f$ electron. Therefore, $f$ electron posses the dual nature of local and itinerant even at very low temperatures, which makes the system complex and difficult to understand [11, 12].

HF compounds Ce$_m$M$_n$In$_{3m+2n}$ ($M$ = Co, Rh, Ir, and Pt) are very suitable for investigating dimensionality and hybridization, and a lot of studies have been carried out on such system. Theoretical calculation suggested that Ce 4$f$ electron spectral weight is mainly located above $E_F$, and only a tiny fraction appears below $E_F$ [13]. In recent years, with the improvement of energy and momentum resolution of angle-resolved photoemission spectroscopy (ARPES), it is possible to study the fine structure of heavy quasiparticle bands, such as $c$-$f$ hybridization and the crystal electric field splitting of the spin-orbital split state $f^1_{5/2}$ and $f^1_{7/2}$ [13–23]. The $f^1_{5/2}$ peak observed near $E_F$ is actually the tail of the Kondo resonance peak, and $f^1_{7/2}$ is the satellite peak. According to the single impurity Anderson model (SIAM) [24], generally, the more localized the Ce 4$f$ is, the stronger the $f^0$ peak is, and the weaker the $c$-$f$ hybridization is [25, 26].

In this study, we carried out high-resolution on-resonance ARPES experiments on HF compounds Ce$_m$M$_n$In$_{3m+2n}$, with $m$ = 1, 2, $n$ = 0 − 2, and discuss the localization or itinerancy of Ce 4$f$ in these compounds and the relationship between the dimensionality and the $c$-$f$ hybridization strength. We find that com-pared with the $f^1$-$f^0$ intensity ratio, the $f^1_{5/2}$-$f^0$ intensity ratio can better reflect $c$-$f$ hybridization strength.

TABLE I. Physical properties of Cerium compounds studied in the present study.

| Compounds | Structure | $\gamma$ [mJ/(mol-Ce·K$^2$)] | Order | $T_c$ (K) | $a$ (Å) | $c$ (Å) | $c/a$ | $d_{Ce-M}$ (Å) |
|---|---|---|---|---|---|---|---|---|
| CeIn$_3$ | cubic | 120 [27] | AF | 0.2 @2.6 GPa [4] | 4.6876 | 4.6876 [28] | 1 | - |
| Ce$_2$IrIn$_8$ | tetragonal | 700 [29] | - | - | 4.6897 | 12.195 [30] | 2.600 | 3.73 |
| Ce$_2$RhIn$_8$ | tetragonal | $\approx$400 [29] | AF | 2 @2.3 GPa [31] | 4.667 | 12.247 [30] | 2.624 | 3.74 |
| CeCoIn$_5$ | tetragonal | 290[32] | - | 2.3 [32] | 4.601 | 7.540 [33] | 1.639 | 3.77 |
| CeIrIn$_5$ | tetragonal | 720 [34] | - | 0.4 [34] | 4.668 | 7.515 [34] | 1.610 | 3.76 |
| CeRhIn$_5$ | tetragonal | $\approx$400 [29] | AF | 0.1 [35] | 4.652 | 7.542 [29] | 1.621 | 3.77 |
| CePt$_2$In$_7$ | tetragonal | 328 [36] | AF | 2.1 @3.5 GPa [37] | 4.6093 | 21.627 [38] | 4.692 | 4.98 |

## SUMMARY OF STUDIED COMPOUNDS

The HF compounds Ce$_m M_n$In$_{3m+2n}$ possess a relatively higher $T_c$ than other HF materials. Its crystal structure is composed of $m$-layers of CeIn$_3$ separated by $n$-layers of $M$In$_2$ along the $c$ axis of the tetragonal unit cell. This family has rich phase diagrams and is a good platform to investigate the effect of dimensionality, hybridization, magnetism, superconductivity, etc. The basic physical properties of HF compounds investigated in the present study are shown in Table.1. From top to bottom, CeIn$_3$, Ce$_2 M$In$_8$, Ce$M$In$_5$ and Ce$_2$PtIn$_7$, the electronic structure becomes more two-dimensional, and the superconducting transition temperature $T_c$ is enhanced. The last column $d$ is the Ce-$M$ bond distance, which is the distance between Ce and its nearest neighbors $M$ atoms. It is worth noting that Ce$_2 M$In$_8$ has only one nearest and four next-nearest neighbors $M$ atoms, but Ce$M$In$_5$ has two nearest and eight next-nearest neighbor $M$ atoms. In other words, the hybridization strength of Ce$M$In$_5$ will be significantly stronger than that of Ce$_2 M$In$_8$. The table also gives the value of $c/a$, which is often used to describe the structural dimension.

CeIn$_3$ is the parent and fundamental unit of this family and crystallizes in the infinite-layer (cubic, 3D) structure (P$m$-3$m$) [28]. CeIn$_3$ is an antiferromagnet at ambient pressure with predominantly localized moments, and Sommerfeld coefficient $\gamma$ is 120 mJ mol$^{-1}$ K$^{-2}$, undergoes a superconducting transition at $T_c = 0.2$ K under pressure of 2.6 GPa [4]. In previous ARPES experiments, a weak heavy quasiparticle band with an energy dispersion of 4 meV was observed, indicating a weak $c$-$f$ hybridization [17].

Ce$_2$IrIn$_8$ is a tetragonal compound. The observation of a spin-glass state [39], a relatively high Sommerfeld coefficient ($\approx$ 700 mJ mol$^{-1}$ K$^{-2}$) [29], and absence of long-range magnetic order indicate that the Ce 4$f$ electrons in Ce$_2$IrIn$_8$ have itinerant character. ARPES observed a flat band near the $E_F$ around the $\Gamma$ point, which shows no large difference in Fermi momentum [18].

Ce$_2$RhIn$_8$ with a tetragonal crystal structure is an antiferromagnetic superconductor with two magnetic transitions at $T_N = 2.8$ and 1.65 K [31, 40]. ARPES measurements suggested that the Ce 4$f$ electrons are essentially localized [26, 41, 42].

CeCoIn$_5$ is a well-studied compound. Its low energy behavior is similar to that of the underdoped cuprates, with a relatively high $T_c \approx 2.3$ K at ambient pressure. In the beginning, $f$ electrons were considered to be itinerant at low temperature because the experimental results of optical conductivity [43], de Haas-van Alphen (dHvA) [44, 45], and scanning tunneling microscope (STM) [46, 47] were in good agreement with the theoretical calculation based on fully itinerant $f$ electrons [7, 44]. Earlier ARPES experiments suggested that $f$ electrons were itinerant up to 105 K [22], while subsequent there was evidence to support the dominance of localized $f$ electrons at low temperature [48, 49], which showed an itinerant nature due to partial hybridizations [49]. Recent ARPES experiments suggest that the localized-to-itinerant transition occurs at surprisingly high temperatures, and $f$ electrons are largely localized even at the lowest temperature [19].

CeIrIn$_5$ is a superconductor with $T_c = 0.4$ K at ambient pressure [34]. Early dHvA experiments suggested that $f$ electron delocalizes and involves forming Fermi surface at low temperature, which means that $f$ electrons have itinerant character [44, 45, 50]. Earlier ARPES experiments suggested that $f$ electrons were nearly localized [25]. Later, on-resonance ARPES experiments have observed heavy quasiparticle peaks, suggesting that although 4$f$ electrons are mainly localized, a small itinerant component is responsible for superconductivity [51]. Recently, ARPES measurements found that the localized $f$ electrons evolve into the HF state starting from a temperature much higher than the coherence temperature [20].

CeRhIn$_5$, unlike CeCoIn$_5$ and CeIrIn$_5$, is an antiferromagnetic compound with Néel temperature $T_N = 3.8$ K at ambient pressure [52]. In contrast to CeCoIn$_5$ and CeIrIn$_5$, dHvA experiments suggested that the Ce 4$f$ electrons are localized in CeRhIn$_5$ [44, 45, 50]. The theoretical calculation also gives the picture of localized $f$ states [53]. Different ARPES research groups found seemingly conflicting results. Some believed that 4$f$ electrons were nearly localized in the paramagnetic state

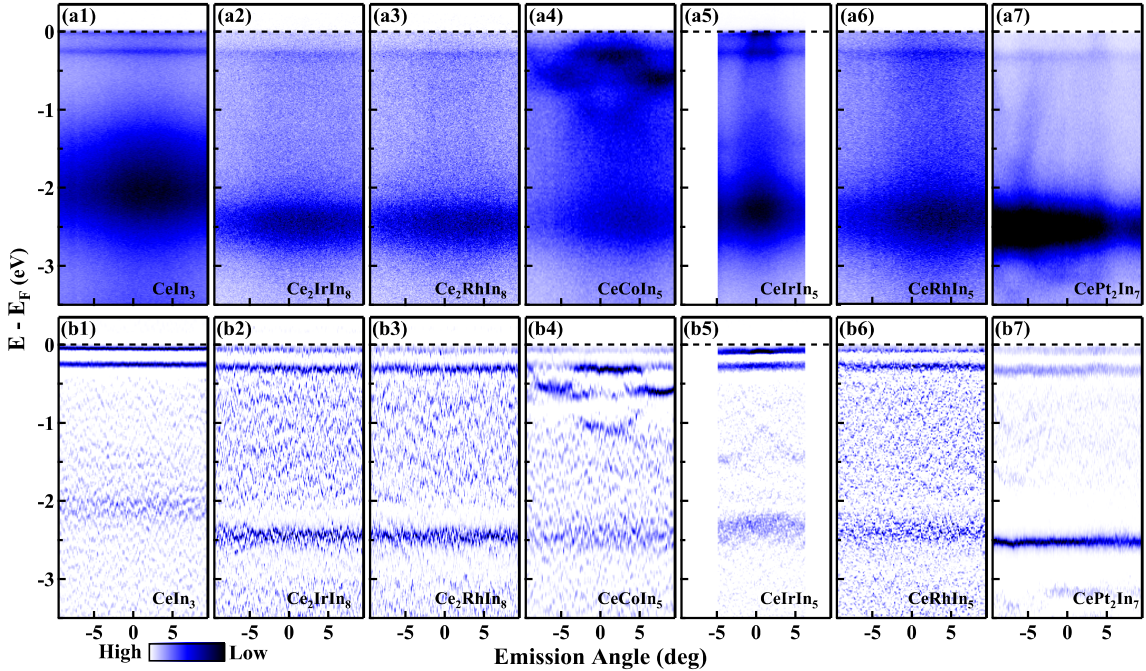

FIG. 1. (color online). Photoemission images for HF compounds $Ce_mM_nIn_{3m+2n}$. **(a1)-(a7)** On-resonance ARPES spectra for $CeIn_3$, $Ce_2IrIn_8$, $Ce_2RhIn_8$, $CeCoIn_5$, $CeIrIn_5$, $CeRhIn_5$, and $CePt_2In_7$, respectively. **(b1)-(b7)** The second-derivative images with respect to energy corresponding to (a1)-(a7), respectively.

[25], while others suggested the $4f$ electrons participate in band formation [54]. Recently, ARPES experiments found band-dependent $c$-$f$ hybridization [21].

$CePt_2In_7$ with body-centered tetragonal crystal structure is an antiferromagnetic superconductor with $T_c = 2.1$ K near 3.5 GPa [37]. Quantum oscillation [55] and ARPES [13, 23, 56] measurements suggested that partially $4f$-electrons contribute to FS formation.

## EXPERIMENTAL DETAILS

High-resolution ARPES experiments were performed on SIS X09LA beamline at the Swiss Light Source, using a VG-SCIENTA R4000 photoelectron spectrometer. The Ce $4f$ electrons characteristics were obtained by on-resonance ARPES at a low temperature of 10 K, which is a practical approach to explore the $4f$ electron states. All compounds were measured with 122 eV photons, except $CeIrIn_5$ and $CePt_2In_7$ compounds were measured with 123 eV photons. The energy and momentum resolution was set to $\sim$ 25-35 meV and 0.2°, respectively. All the samples were single crystals, and cleaved *in situ*, and measured in an ultra-high vacuum.

## RESULTS AND DISCUSSION

Figs. 1 (a1)-(a7) show the spectra taken from freshly cleaved HF compounds $Ce_mM_nIn_{3m+2n}$. Because ARPES spectra is a surface sensitive technique with 122/123 eV photon energies, this results in a large Ce $4f$ spectra weight distribution at deeper binding energies. Figs. 1 (b1)-(b7) display the dispersion reproduced by second-derivative with respect to the energy to sharpen the band structures while maintaining the main band structure. Three flat bands, $f^0$, $f^1_{5/2}$, and $f^1_{7/2}$ final states, can be resolved. The high-intensity flat band observed in all samples corresponds to the $f^0$ state prevalent in the Ce-based HFs material [13–21, 25]. It can be observed that the location, intensity, and width of $f_0$ state vary greatly with materials. Weak heavy quasiparticle bands close to $E_F$ are observed, corresponding to $f^1_{5/2}$ and $f^1_{7/2}$ final states [13–21]. The intensities of $f^1_{5/2}$ and $f^1_{7/2}$ final states also vary significantly with materials. The $f^1_{5/2}$ final state, the tails of Kondo resonant peaks, can be easily distinguished from the raw images of $CeIn_3$ and $CeMIn_5$, while that of $Ce_2IrIn_8$, $Ce_2RhIn_8$ and $CePt_2In_7$ shows only limited spectral weight. And the $f^1_{7/2}$ final state, i.e., the spin-orbit coupling sideband of the $f^1_{5/2}$ state, located at around -300 meV, can be observed in the raw data of all samples. As seen in Fig. 1(a4), the $f^1$ state of $CeCoIn_5$ is heavily overlapped by Co $3d$ state, giving the fact that the photoionization cross-section of the Co $3d$ level is in one order higher than that of the Rh $4d$ and Ir $5d$ at the Ce $4d$-$4f$ excitation threshold.

Figure 2 displays the energy distribution curves (EDCs) of $Ce_mM_nIn_{3m+2n}$, which is integrated over the momentum range shown in Fig. 1. The $f^0$, $f^1_{7/2}$ and $f^1_{5/2}$ final states mentioned above were observed in EDCs. These spectral features were widely observed in Ce-based HFs. Several notable features can be seen from the figures. The relative intensities of the three final states change significantly with the material. Only in $CeIn_3$ and $CeIrIn_5$ can sharp peaks of $f^1_{5/2}$ and $f^1_{7/2}$ be observed. For others, the intensity of $f^1_{5/2}$ is considerably weaker than that of $f^1_{7/2}$. The $f^0$ final state intensity is very strong for the whole family members, except $CeCoIn_5$, suggesting that localized electrons dominate the Ce $4f$ states. The $f^0$ intensities of $CePt_2In_7$ and $CeCoIn_5$ were the strongest and the weakest in the family, respectively. According to the SIAM, they perhaps correspond to the most localized (weakest hybridized) and the most itinerant (most hybridized), respectively. The deduction for $CePt_2In_7$ seems to be consistent with the fact that it has the largest $d$ value in this family [37]. However, the inference for $CeCoIn_5$ is inconsistent with previous experimental and calculation results, which suggested that the c-f hybridization weakens in the order of $CeIrIn_5$, $CeCoIn_5$, and $CeRhIn_5$ [7, 21]. It seems thus not a good idea to use the $f^0$ strength alone to reflect the hybridization strength and itinerancy of $f$ electrons. A more suitable parameter is needed to quantitatively reflect the hybridization strength.

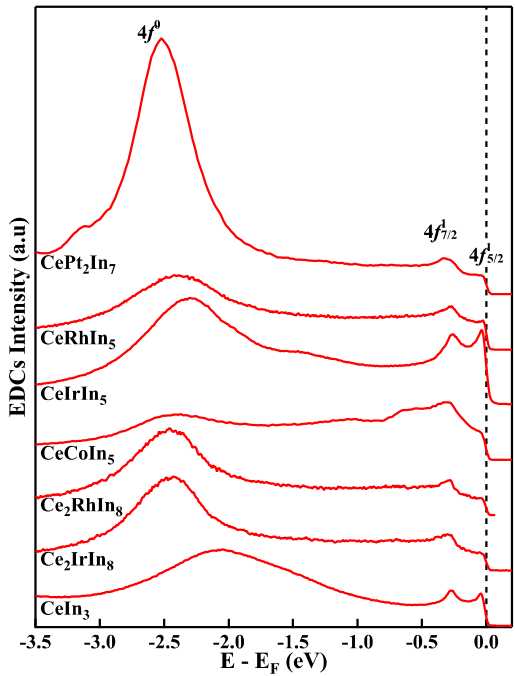

FIG. 2. Comparison of angle-integrated PES spectra of $CeIn_3$, $Ce_2IrIn_8$, $Ce_2RhIn_8$, $CeCoIn_5$, $CeIrIn_5$, $CeRhIn_5$, and $CePt_2In_7$.

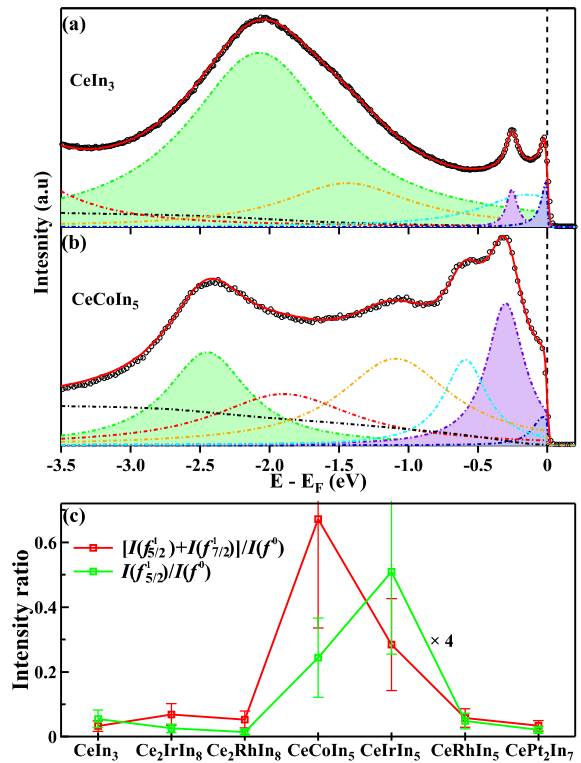

FIG. 3. (a) and (b) Angle-integrated PES Spectra (open circles) and corresponding fitting results (solid lines and dashed dot lines). Filled green, purple, and blue profiles represent the fitting results of $f^0$, $f^1_{7/2}$, and $f^1_{5/2}$ final states, respectively. Other dash-dotted lines originate from conduction bands and background. (c) The $f^1$-$f^0$ and $f^1_{5/2}$-$f^0$ intensity ratios of $Ce_mM_nIn_{3m+2n}$.

The $4d$-$4f$ on-resonance spectra in Ce-based HF compounds can be well understood by the SIAM [24–26], which considers that the ground state is a linear combination of the $f^0$ and $f^1_{5/2}$ states. Previous experiments suggested that the stronger the c-f hybridization, the stronger the $f^1$ peaks [25, 26]. To compare hybridization strength quantitatively, the $f^1$-$f^0$ intensity ratios $[(I(f^1_{7/2})+I(f^1_{5/2}))/I(f^0)]$ were calculated, which has been considered to reflect the c-f hybridization strength [25]. The angle-integrated PES spectra were well fitted with multiple Lorentzian peaks [16]. Figs. 3(a) and (b) display the typical fitted results to $CeIn_3$ and $CeCoIn_5$ spectra. The filled shadow curves represent the heavy $f$ responses. And as evidenced by red, yellow, and cyan dash-dotted lines, the spectra include a large contribution of conduction bands.

Fig. 3(c) display the calculated $f^1$-$f^0$ intensity ratio, together with $f^1_{5/2}$-$f^0$ intensity ratio $I(f^1_{5/2})/I(f^0)$. It can be seen that both intensity ratios $f^1$-$f^0$ and $f^1_{5/2}$-$f^0$, vary greatly with the materials. From $CeIn_3$ to $CePt_2In_7$, the ratios do not change monotonically with $c/a$ or $d$. And interestingly, $CeIn_3$, which has a 3D structure, and $CePt_2In_7$, which is towards the two-dimension limit in the $Ce_mM_nIn_{3m+2n}$ family, both have very low intensity ratio and possess the weakest hybridization/itinerant. And the ratios of four antiferromagnets (CeIn$_3$, Ce$_2$RhIn$_8$, CeRhIn$_5$, and CePt$_2$In$_7$) are very small. The strong localized nature of $f$ electron may be responsible for the antiferromagnetic transition at low temperatures. The $f^1$-$f^0$ value of Ce$M$In$_5$ is the largest of this family, suggesting Ce$M$In$_5$ may have the strongest $c$-$f$ hybridization strength in the Ce$_m M_n$In$_{3m+2n}$ family.

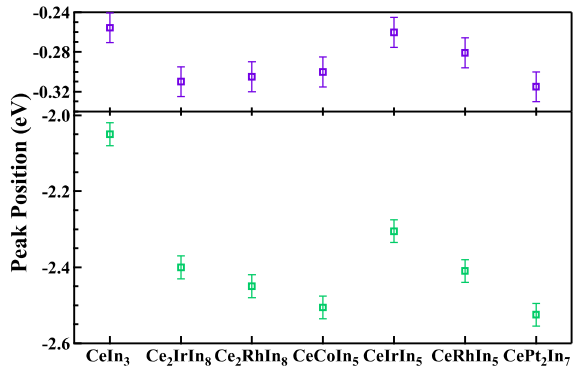

FIG. 4. Energy positions of $f^0$ and $f^1_{7/2}$ final states.

However, in the subfamily Ce$M$In$_5$, the $f^1$-$f^0$ intensity ratio decreases in the order of CeCoIn$_5$, CeIrIn$_5$, and CeRhIn$_5$. According to the SIAM, the $c$-$f$ hybridization weakens in the same order, which is however inconsistent with the previous experimental and theoretical results [7, 21]. But, the $f^1_{5/2}$-$f^0$ intensity ratio is in line with previous results [7, 21], which seems to reflect the hybridization strength of Ce$M$In$_5$ better. Both intensity ratios could thus be employed, but our results suggest that the $f^1_{5/2}$-$f^0$ intensity ratio is more suitable than the $f^1$-$f^0$ to quantify the hybridization strength. We note that the change of atomic distance $d$ in subfamily Ce$M$In$_5$ is very small, but the hybridization strength changes significantly. Similarly, the change of $c/a$ in Ce$M$In$_5$ is not consistent with the change of hybridization strength. However, the $c/a$ ratio can roughly distinguish the hybridization strength among different subfamilies. If the number of adjacent $M$ atoms is taken into account, $d$ can also be used to distinguish the hybridization strength among different subfamilies.

Finally, Fig. 4 summarizes the energy location of $f^1_{7/2}$ and $f^0$ final state of Ce$_m M_n$In$_{3m+2n}$. Since the $f^1_{5/2}$ state close to $E_F$ is the tail of Kondo resonant peak, it is greatly influenced by the resolution-convoluted Fermi-Dirac distribution. The $f^1_{5/2}$ will not be discussed here. The energy locations of $f^1_{7/2}$ and $f^0$ final states vary significantly with the material. They have similar behaviors with the change of materials, but not at the same magnitude. The variation of $f^0$ position is nearly an order of magnitude larger than that of $f^1_{7/2}$. For example, the position of $f^1_{7/2}$ and $f^0$ peaks are -0.26 and -2.08 eV at CeIn$_3$ and -0.32 and -2.53 eV at CePt$_2$In$_7$. The $f^1_{7/2}$ and $f^0$ states are shifted down by 0.06 and 0.45 eV, respectively.

## CONCLUSION

We have studied the electronic structures of HF compounds Ce$_m M_n$In$_{3m+2n}$ by ARPES and observed the heavy quasiparticle bands $f^1_{5/2}$, $4f^1_{7/2}$ and $f^0$ in all compounds. We find that the strength of all three peaks varies greatly with materials composition. The hybridization strength of the subfamily Ce$M$In$_5$ is the strongest of the Ce$_m M_n$In$_{3m+2n}$ family. Compared with the $f^1$-$f^0$ intensity ratio, the $f^1_{5/2}$-$f^0$ intensity ratio is a better indicator of hybridization strength. The $c/a$ ratio and Ce-$M$ band distance $d$, considering the number of adjacent $M$ atoms, set the overall hybridization strength within each subfamily.

## ACKNOWLEDGMENTS

This work was supported by the National Natural Science Foundation of China (Grant No. 12074436, 11574402), and the Innovation-driven Plan in Central South University (Grant No. 2016CXS032). J.R. and P.M.O. acknowledge support through the Swedish Research Council (VR) and the Swedish National Infrastructure for Computing (SNIC), for computing time on computer cluster Triolith at the NSC center Linköping (supported by VR Grant No. 2018-05973). Y.S. acknowledges the support from the Swedish Research Council (VR) through a Starting Grant (Dnr. 2017-05078). O.T. acknowledges support from the Swedish Research Council (VR) and the Knut and Alice Wallenberg foundation. M.M. is partly supported by a Marie Sklodowska-Curie Action, International Career Grant through the European Commission and Swedish Research Council (VR), Grant No. INCA-2014-6426, as well as a VR neutron project grant (BIFROST, Dnr. 2016-06955). Further support was also granted by the Carl Tryggers Foundation for Scientific Research (Grant No. CTS-16:324 and No. CTS-17:325). Work at Los Alamos was performed under the auspices of the U.S. Department of Energy, Office of Basic Energy Sciences, Division of Materials Sciences and Engineering.

---